\documentclass[aps,prb,floatfix,superscriptaddress,reprint]{revtex4-2}
\usepackage{eurosym}
\usepackage{amsbsy}
\usepackage{array}
\usepackage{pifont}
\usepackage{latexsym,epsfig,graphicx}
\usepackage{dcolumn}
\usepackage{subfigure}
\usepackage{comment}
\usepackage{multirow}
\usepackage{color}
\usepackage{bm}
\usepackage{mathrsfs}
\usepackage{amssymb}
\usepackage{amsfonts}
\usepackage{amsmath}
\usepackage{xspace}
\usepackage{float}
\usepackage{epstopdf}
\usepackage{tabularx}
\usepackage{longtable}
\usepackage[colorlinks=true, letterpaper=true, pdfstartview=FitV, linkcolor=blue, citecolor=blue, urlcolor=blue]{hyperref}
\usepackage[normalem]{ulem}
\usepackage[version=4]{mhchem}
\usepackage{epstopdf}
\usepackage{CJKutf8}

\begin{document}
\title{Decoding Equilibrium and Dynamical Criticality in the 2D Topological Order}
\author{Xiao-Ming Zhao(\begin{CJK*}{UTF8}{gbsn}赵小明\end{CJK*})}
\affiliation{\mbox{Department of Physics, University of Science and Technology Beijing, Beijing 100083, China}}
\affiliation{{Key Laboratory of Multiscale Spin Physics (Ministry of Education), Beijing Normal University, Beijing 100875, China}}
\author{Cui-Xian Guo(\begin{CJK*}{UTF8}{gbsn}郭翠仙\end{CJK*})}
\affiliation{{Beijing Key Laboratory of Optical Detection Technology for Oil and Gas, China University of Petroleum-Beijing, Beijing 102249, China}}
\affiliation{{Basic Research Center for Energy Inter disciplinary, College of Science, China University of Petroleum-Beijing, Beijing 102249, China}}
\author{Gaoyong Sun(\begin{CJK*}{UTF8}{gbsn}孙高永\end{CJK*})}
\affiliation{College of Physics, Nanjing University of Aeronautics and Astronautics, Nanjing 211106, China}
\affiliation{Key Laboratory of Aerospace Information Materials and Physics (NUAA), MIIT, Nanjing 211106, China}
\author{Su-Peng Kou(\begin{CJK*}{UTF8}{gbsn}寇谡鹏\end{CJK*})}
\email{spkou@bnu.edu.cn}
\affiliation{{Key Laboratory of Multiscale Spin Physics (Ministry of Education), Beijing Normal University, Beijing 100875, China}}
\affiliation{ {Center for Advanced Quantum Studies, School of Physics and Astronomy, Beijing Normal University, Beijing 100875, China}}

\begin{abstract}
Analytically connecting equilibrium criticality and dynamical quantum phase transitions (DQPTs) under complex driving fields remains a significant challenge, primarily due to the combinatorial complexity of non-local long-range entanglement. Here, we decode this connection in the 2D strongly interacting Wen-plaquette model. By mapping its anyonic excitations to 1D effective dissipative channels, we reveal that microscopic single-particle fidelity zeros exactly reconstruct the macroscopic equilibrium topological phase boundaries. Beyond equilibrium, we demonstrate that during non-unitary quench dynamics, these very same static singularities enforce a momentum-space exclusion against dynamical Fisher zeros. Furthermore, a newly identified dissipation-phase racing mechanism prematurely depletes the decaying mode, suppressing DQPTs and generating topologically trivial steady states. Our results establish exact microscopic static singularities as an analytical decoder for macroscopic non-unitary topological dynamics involving discrete symmetry breaking.
\end{abstract}

\maketitle

\noindent \textbf{\textit{Introduction.---}} The precise characterization of quantum phase transitions (QPTs) has consistently driven the frontiers of condensed matter physics. Traditionally, equilibrium QPTs are diagnosed by the ground-state fidelity and its susceptibility \cite{GuSJ2008,Gu2010, ZengChen2024}. Extending these concepts into the time domain led to the discovery of dynamical quantum phase transitions (DQPTs) \cite{Heyl2013, HeylM2018}, marked by non-analytic singularities in the Loschmidt echo during real-time evolution. This phenomenon has been extensively investigated across diverse quantum systems and quench protocols \cite{Sehrawat2021,Banuls2025, Zeng2023b,ZYZheng2026}. Crucially, profound connections between equilibrium and dynamical QPTs have been established and extensively characterized through universal scaling, topological classifications, and out-of-time-ordered correlators \cite{Ding2020,Vajna2015, Zeng2025a, Salasnich2026,Heyl2018, Nie2020, Zhang2011}. Concurrently, the seminal Lee-Yang zero theory, initially formulated for classical statistical mechanics \cite{Yang1952, Lee1952, Biskup2000}, has been successfully extended to the quantum realm by complexifying the driving parameters \cite{Tong2006a, Kist2021, Vecsei2022, Vecsei2023, Li2023, Liu2024, LiuZou2024, Wang2024, Vecsei2025, Li2025_1, Meng2025, He2025, Lv2026, Guo2026}. This analytical continuation elegantly unveils the geometrical structures of quantum critical phenomena through the complex-plane distribution of fidelity zeros, shedding light on Yang-Lee edge singularities and related dynamical criticality \cite{LuSunGY2025, ZouCPL2023, Xu2025,ZhangYouLW2025}.

Recently, the fidelity-zero framework has achieved remarkable success in probing parity-symmetry breaking in quantum Ising models \cite{Sun2026_1, Gu2026} and topological transitions in two-band models \cite{Chen2026_1}. However, applying this framework to strongly interacting topological orders with long-range entanglement presents considerable analytical challenges. Specifically, the extreme combinatorial complexity of the exponentially large Hilbert space, inextricably coupled with the non-local string-net nature of topological excitations, renders the exact analytical calculation of macroscopic many-body overlaps highly intractable. Moreover, bridging these static critical diagnostics to non-unitary quench dynamics remains highly non-trivial. Under complex driving fields, the non-conservation of probability and the non-orthogonality of eigenstates severely invalidate the standard definitions of fidelity and the Loschmidt echo. While recent advancements utilizing biorthogonal associated-state formalisms have successfully captured non-unitary topological jumps in simplified models \cite{Hu2024_1, Sun2021_2}, as well as anatomized non-Hermitian DQPTs in quantum walks and broader theoretical contexts \cite{ZhangXueP2025,FuGXL2025}, it remains elusive how equilibrium fidelity zeros in complex parameter spaces govern the non-equilibrium dynamical singular boundaries in a fully interacting topological system.

To address this gap, we investigate the $\mathbb{Z}_2$ topological order under a complex transverse field via the exactly solvable Wen-plaquette model\cite{Yu2008a, Zhang2011,Kou2009, Wu2012,Dusue2011, Schulz2012, Tupitsyn2010, Yu2013b}. By mapping its directional anyonic excitations to 1D effective dissipative channels, we reveal that microscopic single-particle fidelity zeros reconstruct the macroscopic equilibrium topological phase boundaries. While macroscopic many-body fidelity zeros govern criticality in generic non-integrable systems involving discrete symmetry breaking, this model's integrability allows us to transparently decompose these global singularities into independent single-particle zeros without finite-size artifacts. Extending to non-unitary quench dynamics via the biorthogonal associated-state formulation, we prove that these static singularities enforce an absolute momentum-space exclusion against dynamical Fisher zeros. Furthermore, a dissipation-phase racing mechanism prematurely depletes the decaying mode, annihilating DQPTs and generating topologically trivial steady states. Thus, we establish static geometric singularities as an analytical decoder for non-unitary dynamical criticality.

\begin{figure}[tbp]
  \centering
\includegraphics[clip,width=0.48\textwidth]{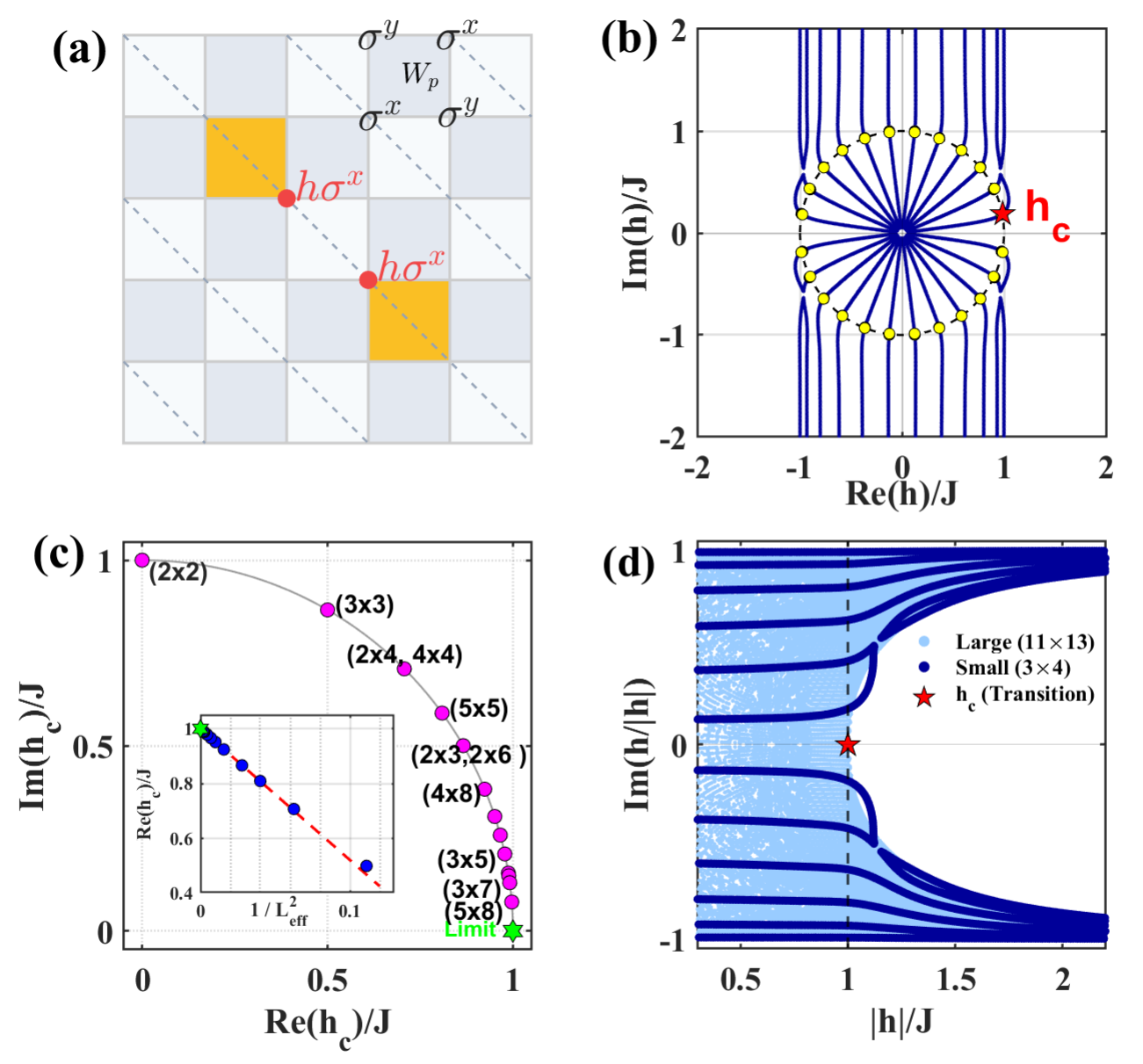}
\caption{Equilibrium characteristics of the 2D Wen-plaquette model. (a) Schematic of the lattice and the constrained quasiparticle motion ($\hat{e}_x - \hat{e}_y$) driven by the transverse field. (b) Distribution of inter-sector many-body fidelity zeros in the complex plane for a finite $3 \times 4$ lattice. The yellow dots denote the intersections of the zero curves with the unit circle $|h_{x}/J|=1$, where the point with the maximum real part is identified as the critical field $h_c$. (c) Finite-size scaling of the critical point $h_c$ extrapolated to the infinite system size. (d) The non-Hermitian phase diagram for the infinite system size, exhibiting clear Yang-Lee edge singularities.}
  \label{fig:static}
\end{figure}

\vspace{2mm}
\noindent \textbf{\textit{Wen-plaquette model and Biorthogonal Framework.---}} We consider the 2D Wen-plaquette model on an $L_x \times L_y$ torus, driven by a complex transverse magnetic field $h_x$:
\begin{equation}
    H_{\mathrm{2D}} = -J \sum_{p} \hat{W}_p - h_x \sum_{i} \sigma_i^x,
    \label{eq:hamiltonian}
\end{equation}
where $J>0$ characterizes the underlying $\mathbb{Z}_2$ topological coupling, and the plaquette operator is $\hat{W}_p = \sigma_i^x \sigma_{i+\hat{x}}^y \sigma_{i+\hat{x}+\hat{y}}^x \sigma_{i+\hat{y}}^y$. Microscopically, such an effective complex transverse field can be engineered via localized dissipative channels coupled to an environment \cite{YangFei2022,XuePengPRL2024}.

The core physics of this model is governed by its emergent anyonic excitations. In the unperturbed topological ground state, all plaquettes satisfy $\hat{W}_p = +1$. Flipping a plaquette eigenvalue to $\hat{W}_p = -1$ creates topological excitations, identified as $e$ and $m$ anyons on the alternating sublattices. Crucially, when perturbed by the transverse field $h_x \sum_i \sigma_i^x$, the local field operator $\sigma_i^x$ strictly anti-commutes with only two diagonally adjacent plaquettes (where it overlaps with $\sigma^y$). Physically, this implies that the external field acts as an ``anyon pump'' that drives the $e$ and $m$ particles to hop \textit{strictly and exclusively} along the diagonal directions $\hat{e}_x - \hat{e}_y$, as intuitively depicted in Fig.~\ref{fig:static}(a).

Due to this severely constrained one-dimensional anyon diffusion mechanism, the 2D strongly interacting topological Hamiltonian is exactly decoupled into a direct sum of independent 1D effective Ising channels: $H_{\mathrm{2D}} = \sum_{n=1}^{\xi} H_{\mathrm{eff}}^{(n)}$ (explicitly detailed in Appendix \textcolor[rgb]{0.00,0.07,1.00}{A}), allowing us to analytically verify our single-particle decoding paradigm without numerical truncation artifacts. Consequently, the global 2D wave function is the exact tensor product of the 1D chains, $|\Psi_{\mathrm{2D}}\rangle = \bigotimes_{n=1}^\xi |\Psi_{\mathrm{1D}}^{(n)}\rangle$, and the total macroscopic energy scales linearly as $E_{\mathrm{2D}} = \xi E_{\mathrm{1D}}$.

From a macroscopic perspective, these 1D channels correspond to closed non-local string-nets wrapping around the two-dimensional torus. The number of these decoupled strings is strictly dictated by the greatest common divisor of the 2D lattice geometry as $\xi = \gcd(L_x, L_y)$. Thus, the macroscopic effective length of each isolated 1D anyon channel is defined as $N_{\mathrm{eff}} = L_x L_y / \xi$. This establishes a direct geometric quantum modulation effect, directly determining the 1D topological correlation length of the system.

The boundary conditions of these chains are locked to the global 2D many-body fermion parity operator $\hat{P}$. The even parity sector ($\hat{P}=1$) enforces fermionic anti-periodic boundary conditions (APBC) with discrete momenta $k_e = \pm \frac{(2m+1)\pi}{N_{\mathrm{eff}}}$ ($m=0,1,2,...,N_{\mathrm{eff}}/2-1$), while the odd parity sector ($\hat{P}=-1$) enforces Fermionic periodic boundary conditions (PBC) with momenta $k_o \in \{ 0, \pi, \pm \frac{2m\pi}{N_{\mathrm{eff}}} \}$ ($m=1,2,...,N_{\mathrm{eff}}/2-1$).

When the transverse field $h_x$ extends into the complex plane, the effective 1D Hamiltonian becomes manifestly non-Hermitian. In momentum space, it maps to a two-band non-Hermitian fermion model described by Pauli matrices, yielding the complex single-particle dispersion $\epsilon(k)$. Consequently, the right and left eigenstates separate and are determined by the non-Hermitian eigenvalue equations $H_k|\psi_{\pm,k}^R\rangle =\pm\epsilon(k)|\psi_{\pm,k}^R\rangle$ and $H_k^{\dagger}|\psi_{\pm,k}^L\rangle =\pm\epsilon^*(k)|\psi_{\pm,k}^L\rangle$.

To systematically circumvent unphysical norm divergences intrinsic to continuous non-Hermitian Hamiltonian dynamics, we explicitly adopt the biorthogonal associated-state framework \cite{Hu2024_1}. For an arbitrary right state $|\Psi\rangle$, its corresponding biorthogonal associated left state $|\tilde{\Psi}\rangle$ is concisely defined as:
\begin{equation}
   |\Psi\rangle = \sum_{\pm} c_{\pm} |\psi_{\pm}^R\rangle \quad \leftrightarrow \quad |\tilde{\Psi}\rangle = \sum_{\pm} c_{\pm} |\psi_{\pm}^L\rangle.
    \label{eq:associated_state}
\end{equation}
This direct transformation stringently ensures that the modified biorthogonal inner product is rigidly normalized and conserved: $\langle \tilde{\psi}_{m,k}^L | \psi_{n,k}^R \rangle = \delta_{mn}$. This associated basis forms the mathematical foundation for analyzing both static fidelity zeros and dynamical geometric phases.

To investigate quantum phase transitions, we evaluate the macroscopic multi-body energy gap $\Delta E_{\mathrm{2D}} = E_{2D}^{\mathrm{odd}} - E_{2D}^{\mathrm{even}}$. Its analytical behavior depends drastically on whether the complex external field falls within the topological region (inside the unit circle, $|h_x| \le J$) or the trivial region ($|h_x| > J$). As detailed in Appendix \textcolor[rgb]{0.00,0.07,1.00}{A}, the macroscopic energy gap is described by a piecewise function :
\begin{equation}
    \Delta E_{\mathrm{2D}} =
    \begin{cases}
     2 J \xi \sqrt{ \frac{1 - \lambda^2}{\pi N_{\mathrm{eff}}} } \lambda^{N_{\mathrm{eff}}} \left[ 1 + \mathcal{O}\big( N_{\mathrm{eff}}^{-1} \big) \right], & |\lambda| \le 1 \\
     \xi \left(E_{1D}^{\mathrm{ odd}} - E_{1D}^{\mathrm{even}} + \epsilon_{k_{\mathrm{min}}}\right), & |\lambda| > 1
    \end{cases}
    \label{eq:piecewise_gap}
\end{equation}
Here, $\lambda \equiv h_x/J$, $k_{\mathrm{min}} = 0$ on the branch connected to $h_{x} = +J$ and
$k_{\mathrm{min}} = \pi$ on the branch connected to $h_{x} =-J$. For $|\lambda| \le 1$, the expression represents the exact leading-order asymptotic gap, with the saddle-point truncation error $\mathcal{O}(N_{\mathrm{eff}}^{-1})$ safely vanishing in the large-size limit. For $\lambda > 1$, the gap is dominated by the bulk quasiparticle excitation $\epsilon_{k_{\mathrm{min}}}$ \cite{FineSize1987}. Inside the unit circle, the gap originates from macroscopic quantum tunneling between the nearly degenerate even and odd parity ground states. Outside the unit circle, topological degeneracy is lifted, and the macroscopic gap is strictly dominated by the minimal local single-particle excitation difference between the distinct discrete momentum grids of the two sectors.

\begin{figure}[tbp]
  \centering
\includegraphics[clip,width=0.48\textwidth]{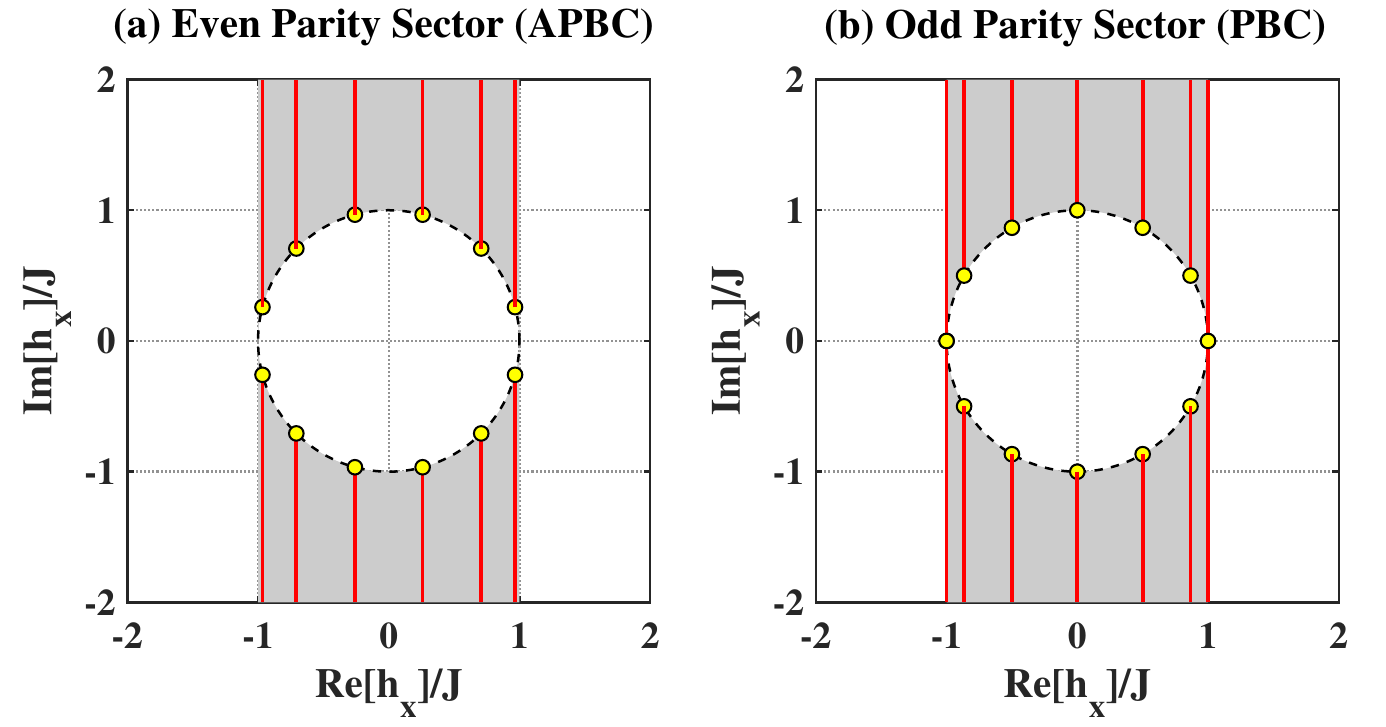}
  \caption{Distributions of intra-sector single-particle fidelity zeros in the complex $h_x/J$ plane for the (a) even (APBC) and (b) odd (PBC) parity sectors. Solid red lines denote the fidelity zeros where $\mathrm{Re}[\epsilon]=0$. The dashed Lee-Yang unit circle ($|h_x|=J$) represents the topological phase boundary, with yellow dots marking the absolute zero-energy points ($\epsilon=0$). The grey background indicates the continuous zero distribution in the infinite-size limit. While finite-size momentum quantization causes the discrete red lines to differ slightly between sectors, they perfectly coalesce into the identical grey-shaded regions as $N_{\mathrm{eff}} \to \infty$.}
    \label{fig:zeros}
\end{figure}

\vspace{2mm}
\noindent \textbf{\textit{Fidelity Zeros and Equilibrium Phase Transitions.---}} Having established the biorthogonal associated states and the piecewise energy gap, we formulate the exact equilibrium quantum criticality. We extend the driving parameter into the complex plane and investigate the generalized ground-state fidelity. Considering an infinitesimal perturbation $h_{x} \to h_{x} + \delta h_{x}$ (with the limit $\delta h_{x} \to 0$), the true predictor of a quantum phase transition is the exact vanishing of the global biorthogonal many-body fidelity, $\mathcal{F}(h_{x}, h'_{x}) = |\langle \tilde{\Psi}_g(h_{x}) | \Psi_g(h'_{x}) \rangle|$.

By virtue of the model's integrability, this macroscopic fidelity can be unified into a concise product over all independent momentum modes in the Brillouin zone:
\begin{equation}
    \mathcal{F}(h_{x}, h'_{x}) = \prod_{k>0} \left|\langle \tilde{\psi}_{g,k}^L(h_{x}) | \psi_{g,k'}^R(h'_{x})\right|,
    \label{eq:unified_fidelity}
\end{equation}
where the indices $k$ and $k'$ denote the allowed discrete momenta in the parity sectors of the reference ground state at $h_{x}$ and the perturbed ground state at $h'_{x}$, respectively. Depending on the relative location of $h_{x}$ and $h'_{x}$ in the complex parameter plane, many-body fidelity zeros emerge through two distinct scenarios based on parity sector occupancy.

\begin{figure*}[tbp]
  \centering
\includegraphics[clip,width=0.99\textwidth]{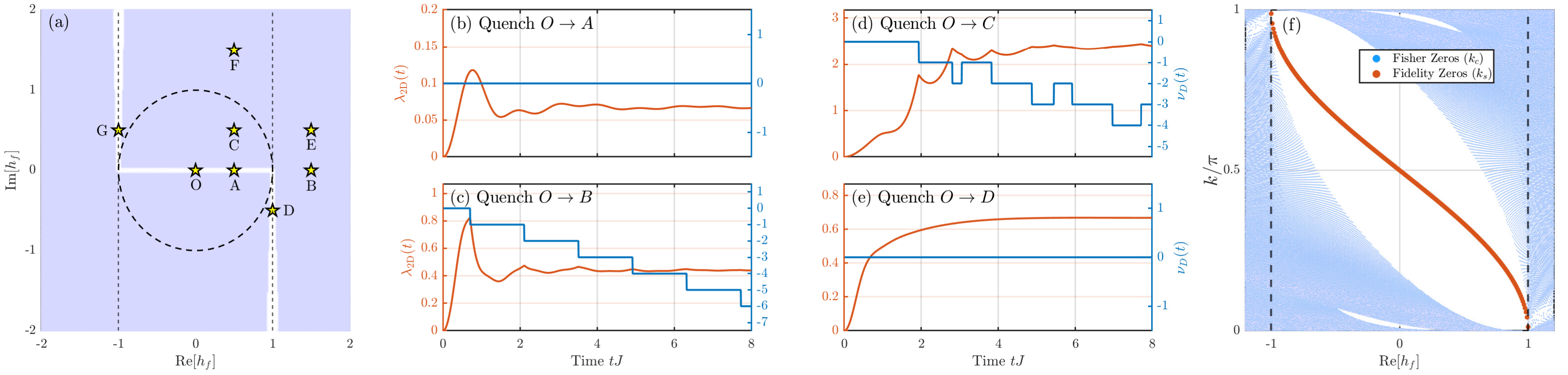}
  \caption{Global steady-state dynamical topological phase diagram and non-Hermitian quench dynamics in the Wen-plaquette model. (a) Scatter plot of the asymptotic DTOP $\nu_D(\infty)$ in the complex quenching field space $(h_R, h_I)$. The initial state is fixed at the Hermitian topological center $O(0,0)$. Blue (white) regions indicate topologically nontrivial (trivial) steady states with $\nu_D \neq 0$ ($\nu_D = 0$). The horizontal white band strictly highlights the Hermitian limits where DQPTs exclusively require crossing the equilibrium quantum critical point ($|h_R/J| > 1$). Representative quenching targets $A$ through $G$ are marked with yellow stars. (b)-(e) Time evolution of the macroscopic return rate $\lambda_{\mathrm{2D}}(t)$ (red lines, left axes) and the instantaneous DTOP $\nu_D(t)$ (blue lines, right axes) for quenches from $O$ to targets $A, B, C$, and $D$, respectively. (f) The absolute $k$-space exclusion zone between dynamical Fisher zeros $k_c$ (blue dots) and static fidelity zeros $k_s$ (red dots), under $|h_f/J| \ge 1$.}
  \label{fig:dynamics}
\end{figure*}

The first scenario corresponds to the inter-sector many-body zeros, occurring when the absolute ground state undergoes a macroscopic parity crossover ($h_{x}$ and $h'_{x}$ reside in different sectors). In this case, the momenta $k \in \{k_e\}$ and $k' \in \{k_o\}$ (or vice versa) belong to different discrete grids. Because wavefunctions from orthogonal parity sectors are strictly orthogonal, the overlap in Eq.~(\ref{eq:unified_fidelity}) vanishes whenever the real part of the macroscopic gap closes ($\mathrm{Re}[\Delta E_{\mathrm{2D}}(h_x)] = 0$). This mechanism analytically predicts non-Hermitian topological phase transitions \cite{Sun2026_1}. As illustrated in Fig.~\ref{fig:static}(b), for a finite lattice, these inter-sector zeros form discrete radiating roots anchoring the phase boundary. Tracking the critical field $h_c$ reveals a robust finite-size scaling extrapolation towards the infinite system size [Fig.~\ref{fig:static}(c)], establishing the definitive non-Hermitian phase diagram mapped in Fig.~\ref{fig:static}(d), where the topological phase transition strictly occurs at $|h_{x}|/J = 1$. Assuming a symmetric scaling path $L_x = L_y = L$, the effective chain length scales as $N_{eff} = L$, ensuring a well-defined asymptotic limit. As explicitly derived in Appendix \textcolor[rgb]{0.00,0.07,1.00}{B}, inside the unit circle ($|h_x| \le J$), these macroscopic inter-sector zeros form radiating lines following the asymptotic analytical solution:
\begin{equation}
    \theta_m = \frac{(2m+1)\pi}{2N_{\mathrm{eff}}}.
\end{equation}
Conversely, outside the unit circle ($|h_x| > J$), the inter-sector zeros are governed by finite-size vacuum energy corrections, forming curved contours that asymptotically merge with the single-particle branch cuts.

The second scenario arises when $h$ and $h'$ belong to the strictly identical parity sector (intra-sector zeros), where the momentum indices are identical ($k = k'$), with both restricted to either the even ($\{k_e\}$) or the odd ($\{k_o\}$) discrete grids. In this configuration, the macroscopic fidelity vanishes if and only if at least one local single-particle mode undergoes a branch crossing. The biorthogonal orthogonality condition forces this single-mode overlap to vanish precisely when the real part of its complex energy band intersects ($\mathrm{Re}[\epsilon(k)] = 0$). Solving this condition yields the exact analytical locus for the single-particle fidelity zeros \cite{Chen2026_1}:
\begin{equation}
    h_R = J \cos k_s, \quad |h_I| \ge J |\sin k_s|.
    \label{eq:single_body_zero}
\end{equation}
The specific distributions of these intra-sector single-particle zeros are explicitly visualized as solid red lines in Fig.~\ref{fig:zeros}(a) and Fig.~\ref{fig:zeros}(b) for the even and odd parity sectors, respectively. Equation~(\ref{eq:single_body_zero}) reveals two distinct geometric features. First, these intra-sector single-particle zeros manifest as straight vertical lines strictly parallel to the imaginary axis. Second, their lower bounds mathematically ensure that $|h_x|^2 = h_R^2 + h_I^2 \ge J^2$. Therefore, \textit{no intra-sector single-particle zeros exist inside the dashed Lee-Yang unit circle}; they are exclusively distributed outside it.

It is imperative to distinguish the physical implications of these two mechanisms. Generally, an isolated single-particle band crossing at $k_s$ does \textit{not} necessarily induce the closure of the global multi-body gap $\Delta E_{\mathrm{2D}}$, because the latter requires a collective summation over all occupied modes. Thus, the intra-sector vertical lines and the inter-sector curved contours are geometrically distinct at finite sizes. Nevertheless, a underlying topological connection unites them at the phase boundary. The endpoints of the single-particle zero lines---where the single-particle energy spectrum entirely closes ($\epsilon(k_s)=0$)---are marked by the yellow dots situated exactly on the dashed Lee-Yang unit circle in Fig.~\ref{fig:zeros}. As the effective system size approaches infinity ($N_{\mathrm{eff}} \to \infty$), the discrete momentum grids merge into a continuum, and these microscopic endpoints infinitely densely pack to perfectly reconstruct the macroscopic static Lee-Yang phase boundaries. Concurrently, the discrete red lines coalesce into the grey-shaded regions as illustrated in Fig.~\ref{fig:zeros}, directly connecting the microscopic and macroscopic descriptions of non-Hermitian criticality.

\vspace{2mm}
\noindent \textbf{\textit{Biorthogonal Dynamical Quantum Phase Transitions.---}}
Dynamical quantum phase transitions (DQPTs) extend the concept of equilibrium criticality into the time domain, characterized by non-analytic singularities in the real-time evolution of a quantum system. In non-Hermitian systems, the standard definition of the Loschmidt echo loses its physical validity due to the breakdown of probability conservation and the non-orthogonality of eigenstates. To capture the intrinsic topological dynamics and maintain a valid probabilistic interpretation, we adopt the biorthogonal associate state formalism \cite{Hu2024_1}.

At $t=0$, the 2D system is prepared in the lowest energy state of the finite-size Hamiltonian, denoted as $|\Psi(0)\rangle$, inside the topological phase under an initial real transverse field (designated as point O in Fig.~\ref{fig:dynamics}). A sudden quench to a complex target field is then executed. The system dynamics are governed by the fully normalized biorthogonal Loschmidt echo:
\begin{equation}
    \mathcal{L}(t) = \frac{\langle \tilde{\Psi}(0) | \Psi(t) \rangle \langle \tilde{\Psi}(t) | \Psi(0) \rangle}{\langle \tilde{\Psi}(t) | \Psi(t) \rangle \langle \tilde{\Psi}(0) | \Psi(0) \rangle}.
\end{equation}
Because the 2D lattice exactly decouples into $\xi$ identical and independent 1D effective channels, the macroscopic many-body echo strictly factorizes as:
\begin{equation}
    \mathcal{L}(t) = \left( \prod_{k>0} \frac{|G_k(t)|^2}{\langle \tilde{u}_{-,k}(t) | u_{-,k}(t) \rangle} \right)^\xi,
\end{equation}
where the fundamental complex transition amplitude for the single-particle mode in the lower energy branch ($-\epsilon$) is defined as $G_k(t) \equiv \langle \tilde{u}_{-,k}(0) | u_{-,k}(t) \rangle$, with $|\tilde{u}_{-,k}(0)\rangle$ being the systematically constructed biorthogonal associate state of the initial state.

By projecting the initial state onto the biorthogonal basis of the post-quench Hamiltonian $H_f(k)$, the amplitude can be decomposed into the amplifying (gain) and decaying components:
\begin{equation}
    G_k(t) = P_+(k) e^{-i \epsilon_f(k) t} + P_-(k) e^{i \epsilon_f(k) t},
\end{equation}
where $\epsilon_f(k) = \epsilon_R(k) + i\epsilon_I(k)$ is the complex energy spectrum of the post-quench Hamiltonian, and $P_\pm(k) = \langle \tilde{u}_{-,k}(0) | u_{\pm,k}^R \rangle \langle u_{\pm,k}^L | u_{-,k}(0) \rangle$ represent the complex projection probabilities (with $P_+ + P_- = 1$).

Because the normalizing denominator is a sum of absolute squares and remains strictly positive, the macroscopic dynamical time singularities (Fisher zeros) emerge if and only if the complex amplitude entirely vanishes ($G_{k_c}(t_c) = 0$). This singularity manifests as a non-analytic divergence in the 2D macroscopic return rate, universally scaled by the system size $L_x L_y = \xi N_{\mathrm{eff}}$:
\begin{align}
    \lambda_{\mathrm{2D}}(t) &= -\lim_{N_{\mathrm{eff}} \to \infty} \frac{1}{\xi N_{\mathrm{eff}}} \ln \mathcal{L}(t) \nonumber \\
    &= -\lim_{N_{\mathrm{eff}} \to \infty} \frac{1}{N_{\mathrm{eff}}} \sum_{k>0} \ln \frac{|G_k(t)|^2}{\langle \tilde{u}_{-,k}(t) | u_{-,k}(t) \rangle}.
\end{align}

Simultaneously, the non-equilibrium topological characteristics are rigorously quantified by the biorthogonal dynamical topological order parameter (DTOP),
\begin{equation}
    \nu_D(t) = \frac{1}{2\pi} \int_0^\pi \partial_k \phi_{g}(k, t) dk,
\end{equation}
where $\phi_{g}$ is the pure geometric phase isolated from the time evolution (detailed derivations are provided in Appendices \textcolor[rgb]{0.00,0.07,1.00}{C} and \textcolor[rgb]{0.00,0.07,1.00}{E}).

Total destructive interference ($G_{k_c}(t_c) = 0$) necessitates the simultaneous fulfillment of magnitude matching and phase inversion between the two evolving eigenmodes. By decoupling the real and imaginary components, the critical time $t_c$ and the critical momentum mode $k_c$ are determined by the exact transcendental coupled equations (Appendix \textcolor[rgb]{0.00,0.07,1.00}{D}):
\begin{equation}
    \frac{(2n+1)\pi}{2\epsilon_R(k_c)} = -\frac{1}{2\epsilon_I(k_c)} \ln \left( \frac{P_+(k_c)}{P_-(k_c)} \right), \quad n \in \mathbb{Z}.
    \label{eq:tc_mag}
\end{equation}
For comprehensive numerical tracking of all potential Fisher zeros, the branch index is evaluated within an extensive range $n \in [-30, 30]$. Based on this exact analytical framework, we elucidate the fundamental properties of non-Hermitian quench dynamics depicted in Fig.~\ref{fig:dynamics}.

The quench dynamics exhibit distinct topological behaviors depending on the target parameters [Fig.~\ref{fig:dynamics}(b-e)]. In the Hermitian limit, quenching within the topological phase (point A) yields no DQPTs ($\nu_D=0$), whereas crossing the critical point (point B) triggers periodic DQPTs with strictly monotonic topological jumps. For complex targets, quenching to point C induces transient up-and-down DTOP fluctuations, as the non-orthogonal gain and loss modes dynamically compete before the gain mode asymptotically dominates. Conversely, quenching to point D completely suppresses DQPTs ($\nu_D=0$) via a novel dissipation-phase racing mechanism.

Extending this evolution to the infinite-time limit, the asymptotic steady-state DTOP $\nu_D(\infty)$ exactly maps the global phase diagram [Fig.~\ref{fig:dynamics}(a)]. The topologically trivial zones ($\nu_D(\infty) = 0$) stem from two regimes: the Hermitian trivial phase ($|h_R|<1, h_I=0$) and the complex ``dissipation-phase racing'' regions (e.g., point D). In the latter, strong non-Hermitian polarization prematurely depletes the decaying mode before the geometric phase completes its required half-cycle inversion. This temporal mismatch inherently forbids the exact amplitude cancellation necessary to generate Fisher zeros.

Underpinning these dynamical phenomena, the biorthogonal formulation enforces a momentum-space exclusion between dynamical Fisher zeros ($k_c$) and static fidelity zeros ($k_s$) [Fig.~\ref{fig:dynamics}(f)].  At a static fidelity-zero mode $k_{s}$. Hence $G_{k_{s}}(t)=P_{+}e^{\epsilon_{\mathrm{I}}}+P_{-}e^{-\epsilon_{\mathrm{I}}}$for
every finite real time, so the Fisher-zero condition cannot be satisfied
at $k_{s}$ (as rigorously detailed in Appendix \textcolor[rgb]{0.00,0.07,1.00}{F}).

\vspace{2mm}
\noindent \textbf{\textit{Conclusion.---}}
In summary, exact momentum-space factorization in the integrable 2D Wen-plaquette model reveals that microscopic single-particle fidelity zeros govern both equilibrium and dynamical criticality. Beyond acting as static Riemann branch cuts reconstructing macroscopic topological phase boundaries, these microscopic zeros dynamically enforce an absolute momentum-space exclusion against Fisher zeros. Alongside a dissipation-phase racing mechanism, this absolute exclusion prematurely suppresses DQPTs, yielding topologically trivial steady states.

Crucially, although single-particle zero isolation and the momentum-space exclusion are specific features of this two-band integrable construction, the underlying principle—that macroscopic many-body fidelity zeros dictate non-unitary topological dynamics—can extend to non-integrable systems, provided that the ground state undergoes a transition between distinct discrete-symmetry sectors. However, for systems with continuous symmetries, or in cases where no discrete-symmetry sector switching occurs, the direct applicability of this formalism remains unclear and is left for future investigation.

Experimentally, our predictions are accessible: 2D $\mathbb{Z}_2$ states in superconducting qubits or Rydberg atoms, combined with controlled localized dissipation and ancillary-qubit interferometry, enable direct observation of these complex-field dynamics. Looking forward, extending this analytical framework to competing external fields (e.g., concurrent transverse and longitudinal fields), potentially aided by recent operator-algebraic tools \cite{Rouz2025,Alberto2025}, promises to systematically decode complex multicritical phenomena in broader non-integrable topological phases. Broadly, this perspective provides a versatile analytical tool to unveil underlying connections between macroscopic steady-state topologies and non-unitary dynamical criticality across diverse quantum many-body systems.

\begin{acknowledgments}
This work is supported by Guangdong Basic and Applied Basic Research Foundation (Grant No. 2023A1515110081), Open Fund of Key Laboratory of Multiscale Spin Physics (Ministry of Education), Beijing Normal University (Grant No. SPIN2024K01), Fundamental Research Funds for the Central Universities (Grant No. FRF-TP-22-098A1, FRF-IDRY-24-28), National Key R\&D Program of China (Grant No. 2023YFA1406704), National Natural Science Foundation of China (Grant Nos. 12174030, 12405030), and the open research fund of Beijing National Laboratory for Condensed Matter Physics (Grant No. 2025BNLCMPKF021).
\end{acknowledgments}

\end{document}